\documentclass[showpacs,amsmath,amssymb,aps,pre]{revtex4}

\usepackage{graphicx}
\usepackage{dcolumn}
\usepackage{bm}

\begin{document}

\preprint{}

\title{Viscosity and mutual diffusion in strongly asymmetric binary ionic mixtures}

\author{Sorin Bastea}
\email{bastea2@llnl.gov}
\affiliation{Lawrence Livermore National Laboratory, P.O. BOX 808, Livermore, CA 94550}


\begin{abstract}
We present molecular dynamics simulation results for the viscosity and mutual 
diffusion constant of a strongly asymmetric binary ionic mixture (BIM). We compare the 
results with available theoretical models previously tested 
for much smaller asymmetries. For the case of viscosity we propose a new predictive 
framework based on the linear mixing rule, while for mutual diffusion we discuss some 
consistency problems of widely used Boltzmann equation based models.
\end{abstract}


\maketitle

The important advancements that occurred in the last decade in the experimental 
techniques 
involving high-power lasers have generated a renewed interest in the properties 
of dense plasmas in general and their transport properties in particular \cite{rnz95, ccz98, sc02}. 
The experimental capabilities currently available and the ones that are expected 
to become available in the near future \cite{pbk94} promise to further advance the field 
of inertial confinement fusion (ICF) as well as shed new light on long-standing 
astrophysics problems. Many times such experiments either probe, or their outcome is 
strongly dependent on, the behavior of plasma mixtures with 
various degrees of charge and mass asymmetries of the components. 
Such mixtures occur for example in ICF due to the instability 
(e.g. Richtmyer-Meshkov or Rayleigh-Taylor) driven mixing of the heavy elements that 
make up the enclosing shell and the much lighter fuel. In this case 
the stability of the initial interfaces, nature of the ensuing flows and 
degree of fuel 
contamination are crucially linked to such mixture properties as shear 
viscosity and mutual diffusion \cite{rzb03,acr03,mam04}.
In the present paper we calculate these properties using molecular dynamics 
simulations for a simple but relevant plasma model: the binary ionic mixture (BIM) \cite{htv77,hmv79}, 
which is a generalization of the one-component plasma (OCP) \cite{h73}. 
We study a rather extreme, ICF-inspired system \cite{mam04}, $D^{+}-Au^{39+}$, which displays 
roughly two orders of magnitude charge and mass asymmetry,  but the results should apply 
to other plasma mixtures as well, e.g. of astrophysics significance \cite{sch03}, 
where large charge and mass asymmetries are present. We compare the MD results with available 
theoretical models previously tested 
for much smaller asymmetries. For the case of viscosity we propose a new predictive 
framework based on the linear mixing rule, while for mutual diffusion we discuss some 
consistency problems of Boltzmann equation based models.

The BIM model consists of a mixture of $N_1$ point ions of charge $q_1=Z_1e$ and mass $M_1$ 
and $N_2$ point ions of charge $q_2=Z_2e$ and mass $M_2$ embedded in a uniform, rigid, 
neutralizing electronic background. We denote the number concentrations by 
$x_{\alpha}=N_{\alpha}/N$, $N=N_1+N_2$, $\alpha=1,2$ and number densities by 
$\rho_{\alpha}=N_{\alpha}/V$, $\rho=\rho_1+\rho_2$, where $V=L^3$ is the volume. 
$\langle Z \rangle=x_1Z_1+x_2Z_2$ is the average charge and $\rho\prime=\rho_1Z_1+\rho_2Z_2$ the 
electronic number density. As usual the mean inter-electronic and inter-ionic distances are defined 
by $a\prime=(3/4\pi\rho\prime)^{\frac{1}{3}}$ and 
$a=(3/4\pi\rho)^{\frac{1}{3}}=a\prime\langle Z \rangle^{\frac{1}{3}}$, while the electronic and 
ionic coupling parameters are:
\begin{eqnarray}
&&\Gamma^\prime=\frac{e^2}{a\prime k_BT}=\Gamma\langle Z \rangle^{\frac{1}{3}}\\
&&\Gamma_i=\Gamma^\prime\langle Z ^{\frac{5}{3}}\rangle
\end{eqnarray}
with
\begin{eqnarray}
\Gamma=\frac{e^2}{a k_BT}
\end{eqnarray}
The thermodynamics of the system is fully determined by one coupling constant, e.g. 
$\Gamma$, and concentration $x_1$. 
As it is the case for the OCP, the validity range of the BIM model is 
such that the Fermi temperature of the electrons is $T_F\gg T$ and the densities are 
high enough so $r_s\ll 1$, $r_s=a\prime/a_0$, $a_0$ - Bohr radius, corresponding 
to a completely degenerate and rigid electronic background. 

The thermodynamics of the BIM has been thoroughly studied 
and is known to be very well described by the linear mixing rule \cite{htv77,oii93,dsc96,dss90}. 
For moderate charge asymmetries an OCP-based ``one-fluid theory'' is also a reasonable 
approximation \cite{htv77}, with an effective charge 
$Z^2_{eff}=\langle Z^{\frac{5}{3}}\rangle\langle Z\rangle^{\frac{1}{3}}$ 
suggested by the ion-sphere model \cite{ees54}.
The relative success of this ``one-fluid'' representation has lead Clerouin et al. 
to propose that the shear viscosity can also be predicted in terms of the 
equivalent OCP, as already tested for thermal transport \cite{pc90}. 
(Although not explicitly stated in \cite{ccz98}, 
further assumptions need to be made for such a prediction - see below).
This ``one-fluid'' 
approach was shown to be suitable for calculating the BIM viscosity at charge and mass 
asymmetries of order $\approx 10$. Before testing this idea on much larger asymmetries, 
$\approx 100$, we note that, surprisingly, the viscosity of the OCP itself does not 
appear to be very accurately known. For intermediate and strong couplings, $1\leq \Gamma\leq 100$, 
Bernu and Vieillefosse \cite{bv78} 
have proposed an interpolation formula based on three MD simulation results 
obtained with systems of 128 - 250 particles, while at even stronger couplings Ichimaru and Tanaka 
have introduced a generalized viscoelastic theory \cite{it86}. 
In \cite{ccz98} the authors propose a different relation based on the 
kinetic theory of Wallenborn and Baus \cite{wb78}, that extends to the weak coupling regime. 
However, the disagreement between these two approaches (Refs. \cite{bv78} and \cite{ccz98}) 
is significant in the regime that they both cover, $\Gamma \geq 1$, particularly at intermediate $\Gamma$'s. 
Unfortunately it is difficult to ascertain the reliability of these predictions given 
both the limited simulation 
results available and the small system size used, which limits the accuracy of the results. 

To settle this question we performed extensive microcanonical MD simulations of the OCP with much 
larger system sizes - 1372 particles, and a wide range of coupling constants, 
$0.05\leq \Gamma\leq 100$. The Coulomb interactions 
were handled using the Ewald summation technique with conducting boundary conditions. The 
calculation of the shear viscosity $\eta$ was done using the Green-Kubo relation:
\begin{eqnarray}
\eta = \frac{1}{Vk_BT}\int^{\infty}_0\langle 
{\bf \hat{\sigma}}_{xy}(t){\bf \hat{\sigma}}_{xy}(0)\rangle dt
\label{eq:gk}
\end{eqnarray}
As shown by Bernu and Vieillefosse \cite{bv78}, and more recently in the context of 
Yukawa plasmas by Salin and Caillol \cite{sc03}, the evaluation of the pressure tensor 
${\bf \hat{\sigma}}$ requires an Ewald-type summation for its interaction part, ${\bf \hat{\sigma}}^I$:
\begin{eqnarray}
&&{\bf \hat{\sigma}}={\bf \hat{\sigma}}^K+{\bf \hat{\sigma}}^I\\
&&{\bf \hat{\sigma}}^K_{ab}=\sum_i M_i v_{i,a}v_{i,b}\\
&&{\bf \hat{\sigma}}^I={\bf \hat{\sigma}}^{(r)}+{\bf \hat{\sigma}}^{(k)}\\
&&{\bf \hat{\sigma}}^{(r)}_{ab}=\frac{1}{2}\sum_{i\neq j}q_i q_j\frac{r_{ij,a}r_{ij,b}}{r_{ij}}
\left[\frac{2 \alpha e^{-\alpha^2 r^2_{ij}}}{\surd\pi r_{ij}}+\frac{erfc(\alpha r_{ij})}{r^2_{ij}}\right]\\
&&{\bf \hat{\sigma}}^{(k)}_{ab}=\frac{2\pi}{L^3}\sum_{|{\bf k}|\neq 0}
\frac{e^{-\frac{k^2}{4\alpha^2}}}{k^2}\left[\delta_{ab}-2\left(1+\frac{k^2}{4\alpha^2}\right)
\frac{k_ak_b}{k^2}\right]|\tilde\rho({\bf k})|^2\\ 
&&\tilde\rho({\bf k}) = \sum_i q_i e^{-i{\bf k}\cdot{\bf r_i}}
\end{eqnarray}
where $a$ and $b$ denote the Cartesian coordinates. The Ewald parameter $\alpha$ \cite{dh98} 
was chosen such that the real space sums, e.g. 
${\bf \hat{\sigma}}^{(r)}$, can be calculated with the usual minimum-image convention,  
as shown above. The duration of the runs was $10^3-10^4\omega^{-1}_p$, 
where $\omega^2_p=4\pi\rho e^2/M$ is the plasma frequency $(Z=1)$. The natural unit for the viscosity 
of the OCP is $\eta_0=\rho M a^2 \omega_p$.

\begin{figure}
\includegraphics{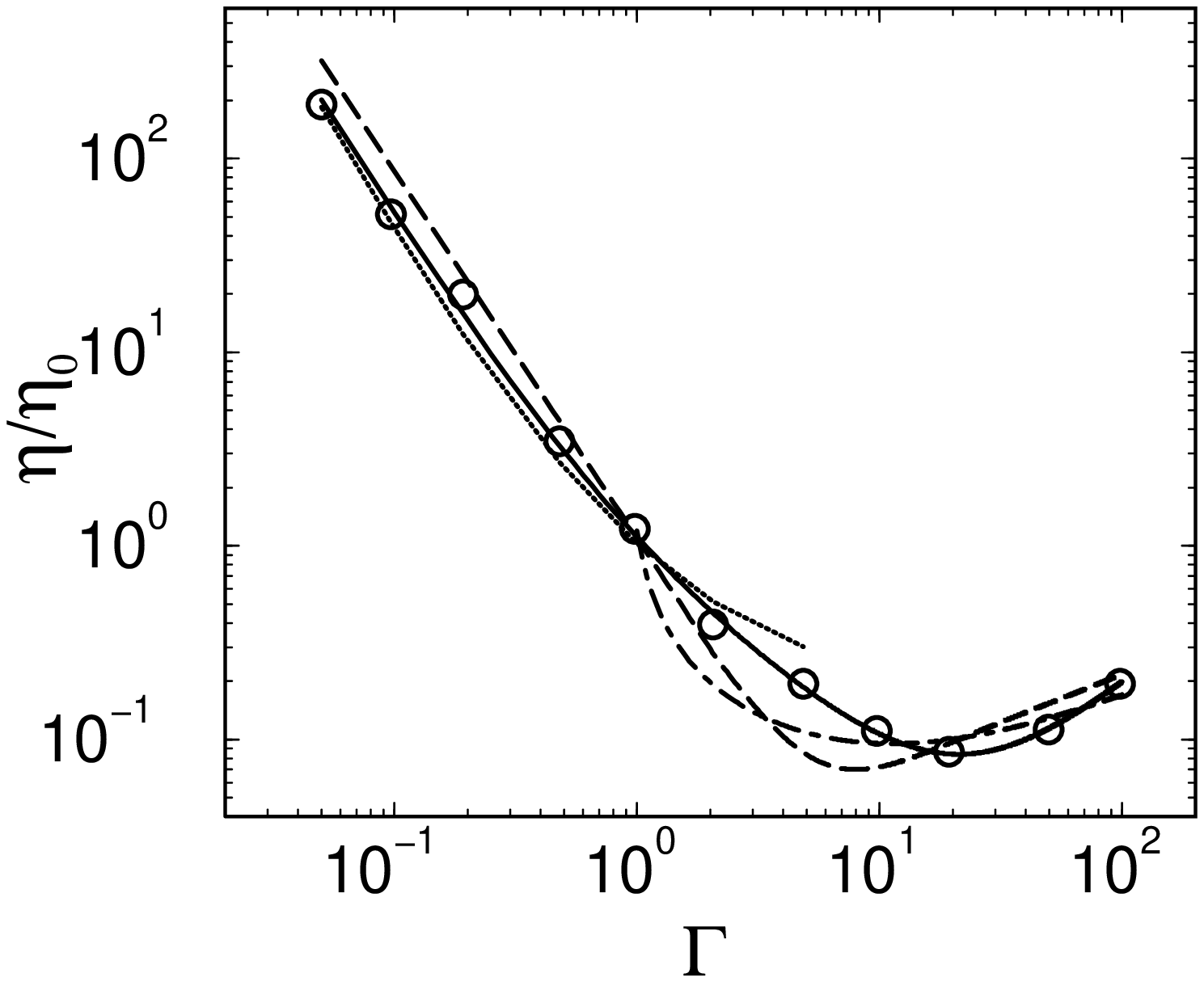}
\caption{OCP viscosity: present simulations (circles), fit of simulation results - Eq. \ref{eq:visocp} 
(solid line), predictive relation of Ref.  \cite{ccz98} (dashed line), 
Bernu-Vieillefosse interpolation formula \cite {bv78} (dot-dashed line), 
Chapman-Enskog estimate using screened potentials (dotted line).}
\label{fig:ocp}
\end{figure}

Our simulation results are shown in Fig. \ref{fig:ocp}, together with the interpolation formula of Bernu and 
Vieillefosse (for $\Gamma\ge 1$) and the relation proposed in \cite{ccz98} based on OCP kinetic theory. 
The errors, estimated using a standard block analysis \cite{altil}, are between about 
$8\%$ at intermediate and large $\Gamma$ and $25\%$ at the lowest $\Gamma$. 
The present viscosity results largely agree with those of \cite{bv78}, but suggest that in the 
intermediate coupling range the OCP viscosity is significantly higher than previously 
predicted, in agreement with \cite{dn00}. The well known viscosity minimum appears to 
be around $\Gamma\simeq 21$, with 
$\eta/\eta_0\simeq 0.084$. We are not aware of other simulations for weakly coupled plasmas, 
$\Gamma\le 1$, but our viscosity results in this regime are in qualitative agreement with 
the kinetic theory of 
Wallenborn and Baus, although somewhat lower. Since for these conditions the screened, Debye-H\"uckel 
potential $e^2exp(-r/\lambda_D)/r$, 
$\lambda_D=\left[k_B T/4\pi e^2 \rho\right]^{\frac{1}{2}}$, should be an appropriate representation 
of the effective inter-ionic interaction and $a/\lambda_D\propto \Gamma^{\frac{1}{2}}$, i.e. the 
interaction can be fairly short-ranged even for $\Gamma$ significantly smaller than $1$, 
a good viscosity estimate should be provided by the 
Chapman-Enskog theory, $\eta=5k_B T/8\Omega^{(2)}_2$ \cite{cc}. As shown in Fig. \ref{fig:ocp} 
(see further below for how the collision integrals $\Omega^{(2)}_2$ are calculated) this is indeed 
the case for $0.05\le\Gamma\le 1$. We choose to fit all the data points with the relation:
\begin{eqnarray}
\frac{\eta}{\eta_0}=A\Gamma^{-2}+B\Gamma^{-s}+C\Gamma
\label{eq:visocp}
\end{eqnarray}
which captures rather well the behavior of the OCP viscosity in the wide range of couplings 
simulated. The best parameters are: $A=0.482$, $B=0.629$, $C=1.88\times 10^{-3}$, and $s=0.878$.

We know tackle the question of the BIM viscosity, in particular when the charge and 
mass asymmetries are very large. It should be noted that the system that we focus 
on, $D^{+}-Au^{39+}$, is not 
a simple BIM {\it per se}, as gold ($Au$) is only partially ionized and the effect of the 
remaining electrons may be important under certain thermodynamic conditions. 
However, the BIM approach 
is still relevant provided the densities and temperatures are such that the distance of 
closest approach between ions is larger that the radius of the remaining ion cores. 

\begin{figure}
\includegraphics{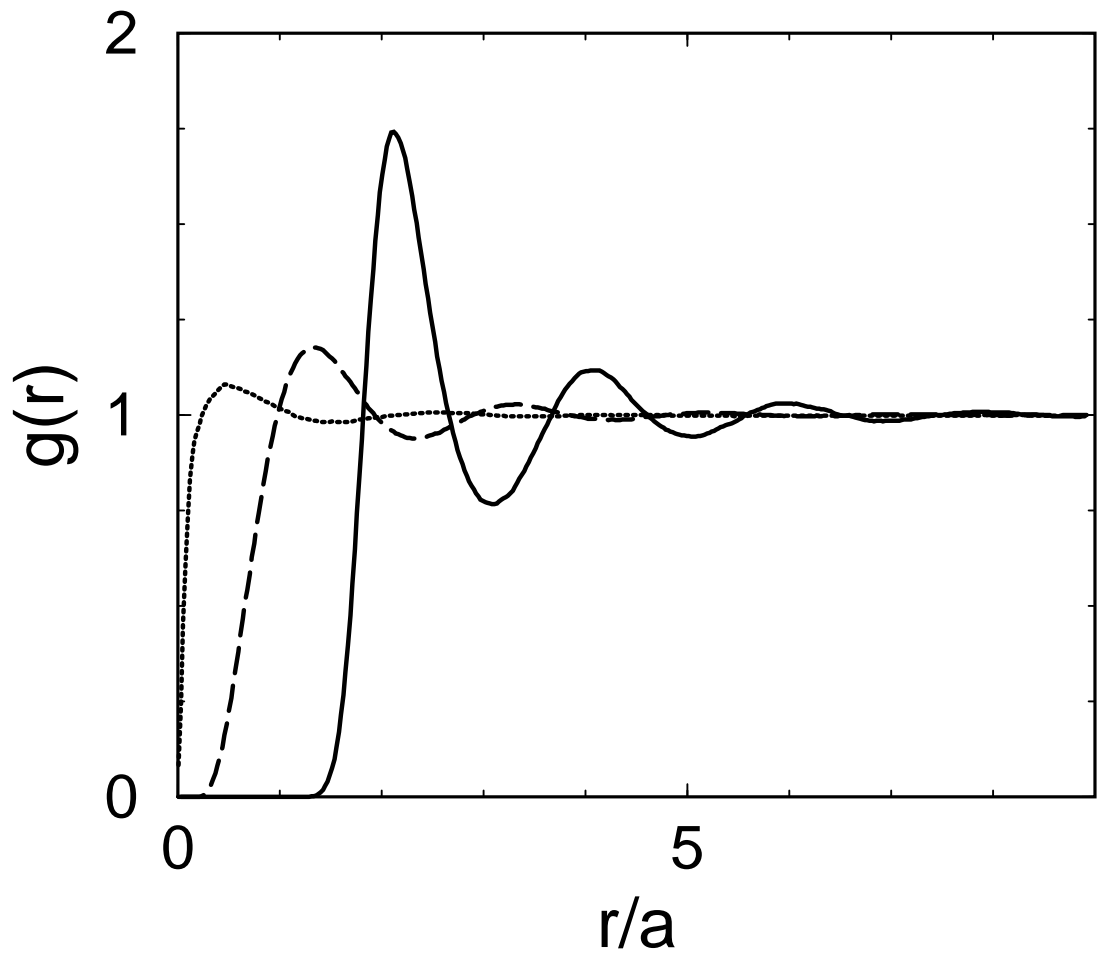}
\caption{Ion-ion pair correlation functions for $\Gamma=0.05$, $x_{Au}=0.5$: $D^+-D^+$ (dotted line), 
$D^+-Au^{39+}$ (dashed line), $Au^{39+}-Au^{39+}$ (solid line).}
\label{fig:gr}
\end{figure}

To elucidate the effect on viscosity of mixing plasmas with such large differences 
in charge and mass as $D^+$ and $Au^{39+}$ it is 
convenient to adopt the procedure of Ref. \cite{ccz98}, where the coupling constant $\Gamma$ is kept 
fixed and the concentration of the two species is varied. We set $\Gamma=0.05$, 
which corresponds to $\Gamma_i=0.05$ for pure deuterium ($x_{Au}=0$), i.e. a weakly coupled plasma, 
and to $\Gamma_i\simeq 76$ for pure gold ($x_{Au}=1$), 
i.e. a strongly coupled plasma. As before, we perform 
microcanonical simulations with a system of 1372 particles, at a number of different 
concentrations $x_{Au}$. Due to the strong charge and mass asymmetries, exceedingly long 
run times are necessary for both equilibration and data accumulation to calculate the 
viscosity using Eq. \ref{eq:gk} with an accuracy of $20-25\%$. A good measure of the large ``size'' 
difference between the ions is provided for example by the pair correlation functions, 
which we show in Fig. \ref{fig:gr} for $x_{Au}=0.5$. 

\begin{figure}
\includegraphics{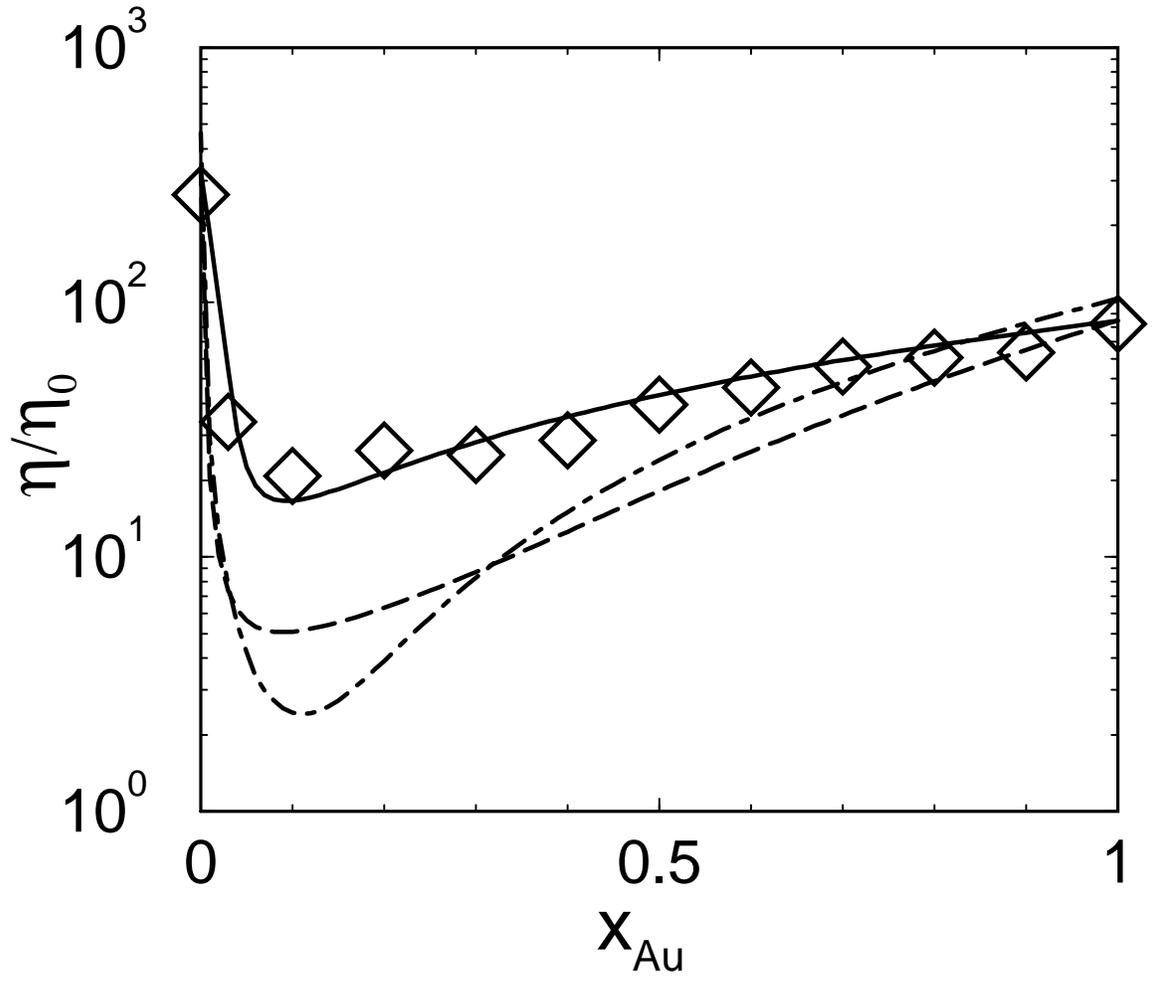}
\caption{Viscosity of the $D^+-Au^{39+}$ BIM at $\Gamma=0.05$ as a function of composition: simulations (diamonds), 
one-component model of 
Ref. \cite{ccz98} (dot-dashed line), one-component model using Eq. \ref{eq:visocp} for the 
OCP viscosity (dashed line), 
two-component model Eq. \ref{eq:vistcp} (solid line).}
\label{fig:vis}
\end{figure}

As noted in \cite{ccz98} the viscosity drops steeply upon mixing highly charged, 
heavy ions in a weakly coupled plasma, see Fig. \ref{fig:vis}. This effect can be understood 
qualitatively in the framework of a ``one-fluid'' theory, which we describe below. 
The coupling constant of the equivalent 
OCP is $\Gamma_{eff}=\Gamma Z^2_{eff}$, where 
$Z^2_{eff}=\langle Z^{\frac{5}{3}}\rangle\langle Z\rangle^{\frac{1}{3}}$ follows from 
the ion-sphere model \cite{ees54}. In this approximation the thermodynamics is 
fully determined by $\Gamma_{eff}$, 
but the calculation of the viscosity requires some 
additional arguments. For example, a reasonable unit for the viscosity of this system may be 
assumed to be $\eta^m_0=\rho\langle M \rangle a^2\omega_{pm}$, $\langle M \rangle=x_1M_1 + x_2M_2$, 
where $\omega^2_{pm}=\omega^2_p\langle Z \rangle^2/\langle M \rangle$ is the ``hydrodynamic'' 
plasma mixture frequency. (The use of the so-called ``kinetic'' mixture frequency \cite{hjm85} leaves the 
results largely unchanged). 
The mixture viscosity in these units is then postulated to be given by the scaled effective-OCP 
viscosity:
\begin{eqnarray}
\frac{\eta(\Gamma, x)}{\eta^m_0(x)}=\frac{\eta_{OCP}(\Gamma_{eff})}{\eta_0}
\end{eqnarray}
We show in Fig. \ref{fig:vis} the results of such calculations for the system that we study, using both the OCP 
viscosity of Ref. \cite{ccz98}, and our prediction Eq. \ref{eq:visocp}. The qualitative dependence of 
the MD results on $x_{Au}$, arising from the opposite behaviors of $\eta_{OCP}(\Gamma_{eff})$ and 
$\eta^m_0(x)$, is reproduced correctly, but the quantitative disagreement is also very significant, 
particularly for small and moderate $Au$ concentrations. The use of the more accurate OCP viscosity Eq. 
\ref{eq:visocp} does not fully alleviate this problem. We conclude that, not surprisingly, 
the accuracy of the ``one-fluid'' model is diminished for extreme asymmetries. 

It is clear that this limitation can only be overcome by taking into account, 
either explicitly or implicitly, 
the mixture asymmetry. A direct calculation along the lines of the kinetic theory of Wallenborn 
and Baus \cite{wb78} has been used for example to determine the BIM mutual diffusion constant at small 
asymmetries \cite{bp87}. 
However, the calculation of the viscosity is even more complex and given the limitations of the theory 
even for the OCP its success for BIM quantitative predictions is rather doubtful.
We turn therefore to a more indirect method, which we outline below.
First, we recall an interesting and much studied property of binary ionic mixtures, 
the linear mixing rule \cite{htv77,oii93,dsc96,dss90}. Hansen et al. have pointed out that the excess internal 
energy of the BIM, $u=U_{ex}/Nk_BT$, is very acurately represented as a linear combination of the 
excess energies of two one-component plasmas with the same electronic coupling constant $\Gamma^\prime$ 
as the BIM (and ionic charges $Z_1e$ and $Z_2e$), i.e. the mixing of the two components at the 
same temperature and electronic density is largely ideal:
\begin{eqnarray}
u(\Gamma^\prime, x_1)\simeq x_1 u_{OCP}(\Gamma^\prime Z^{\frac{5}{3}}_1)+
x_2 u_{OCP}(\Gamma^\prime Z^{\frac{5}{3}}_2)
\end{eqnarray}
We find that this rule is satisfied at very large asymmetries as well, with the largest relative 
deviations occurring at $\Gamma^\prime \ll 1$, i.e. small $Au$ concentrations, in agreement 
with \cite{htv77}. Given this nearly ideal mixing behavior we 
assume that other system properties, e.g. viscosity, are bracketed by the 
component values as well. For an interpolation relation between the viscosities of the two OCP's 
at a given composition we now borrow some concepts from the linear transport theory of 
composite media. A common situation 
encountered in such systems is that of ``impurities'' with a 
generalized conductivity $\alpha_1$ and total volume fraction $\phi_1$ randomly dispersed 
in a matrix $\alpha_2$. Under these circumstances the effective medium theory \cite{rl52} 
employs a mean-field like, self-consistent approximation 
to predict the medium conductivity on scales much larger than those of the inhomogeneities. For 
the case of viscosity this yields  \cite{la01} for the effective medium viscosity $\eta_m$:
\begin{eqnarray}
\sum_i \phi_i\frac{\eta_i - \eta_m}{\eta_i + \frac{3}{2}\eta_m}=0
\label{eq:vistcp}
\end{eqnarray}
We now note that the BIM is obtained by combining 
one-component plasmas with volume fractions $\phi_i=Z_ix_i/\langle Z\rangle$ and 
use the above relation to predict the mixture viscosity, essentially assuming 
that the theory applies for atomically sized ``impurity'' domains. Using Eq. 
\ref{eq:visocp} for the individual OCP viscosities $\eta_1$ and $\eta_2$ 
we obtain the results shown in Fig. \ref{fig:vis}.

The degree of agreement with the simulation results is fairly remarkable. This may lead 
one to believe that the system is perhaps thermodynamically unstable and separating into 
two OCP phases \cite{htv77}. However, we find no evidence for such a scenario and conclude 
that the decoupling signaled by the linear mixing rule along with the tremendous asymmetry 
between the ions lead to behavior mimicking that of a macroscopically mixed system. 

We now turn to the case of ionic interdiffusion in the BIM. Mutual diffusion in plasma mixtures 
plays an important role in the prediction of stellar 
structure \cite{vs98}, as well as the stability of ICF targets \cite{rzb03}. For the case of 
a binary mixture the mutual diffusion coefficient $D_{12}$ can be calculated in terms of the 
fluctuations of the microscopic interdiffusion current \cite{hm}:
\begin{eqnarray}
&&D_{12}=x_1 x_2\left[\frac{\partial^2(\beta G/N)}{\partial x^2_1}\right]_{P,T} D^0_{12}\\
&&D^0_{12}=\frac{1}{3Nx_1 x_2}\int^{\infty}_0\langle {\bf j}_c(t){\bf j}_c(0)\rangle dt\\
&&{\bf j}_c(t)=x_2\sum^{N_1}_1 {\bf v}_i(t)-x_1\sum^{N_2}_1 {\bf v}_i(t)
\end{eqnarray}
where $G$ is the Gibbs free energy.
The thermodynamic prefactor that multiplies the Green-Kubo component $D^0_{12}$ 
reduces to unity for dilute gas mixtures \cite{hm}, but in low density, weakly coupled plasmas 
goes to $\langle Z^2\rangle/\langle Z\rangle^2$, which has been interpreted as an effect of 
the ambipolar electric field of the electrons \cite{bp87}. It is worth noting that the above relation is 
a good estimate for $x_1 x_2 \left[\partial^2(\beta G/N)/\partial x^2_1\right]_{P,T}$ 
at weak as well as strong couplings. 
Since the linear mixing rule holds well for all couplings, the change in Helmholtz free 
energy upon mixing at constant electronic density (and temperature):
\begin{eqnarray}
\frac{\Delta F}{N}=f(\Gamma^\prime, x_1) - x_1 f_{OCP}(\Gamma^\prime Z^{\frac{5}{3}}_1)-
x_2 f_{OCP}(\Gamma^\prime Z^{\frac{5}{3}}_2)
\end{eqnarray}
is very well approximated by the ideal entropy of mixing (with negative sign):
\begin{eqnarray}
\frac{\Delta F}{N}\simeq k_BT\left[x_1 \ln \frac{x_1 Z_1}{\langle Z\rangle} + 
x_2 \ln \frac{x_2 Z_2}{\langle Z\rangle}\right]
\end{eqnarray}
If we assume that the system pressure $p=p_{electronic}+p_{ionic}$ is entirely determined by the 
electronic density, i.e. $p_{electronic}\gg p_{ionic}$, which is consistent with the initial 
assumption of a rigid electronic background, $r_s\ll 1$, then $\Delta G=\Delta F$ \cite{htv77}. 
We can therefore immediately 
calculate the thermodynamic prefactor as $\langle Z^2\rangle/\langle Z\rangle^2$. 
The difference between BIM and ideal gas mixtures appears here to be an entropic 
effect induced by the charge neutralizing background, as mixing occurs at constant 
electronic density (i.e. constant electronic pressure) as opposed to constant molecular density 
(i.e. constant ideal gas pressure). 

\begin{figure}
\includegraphics{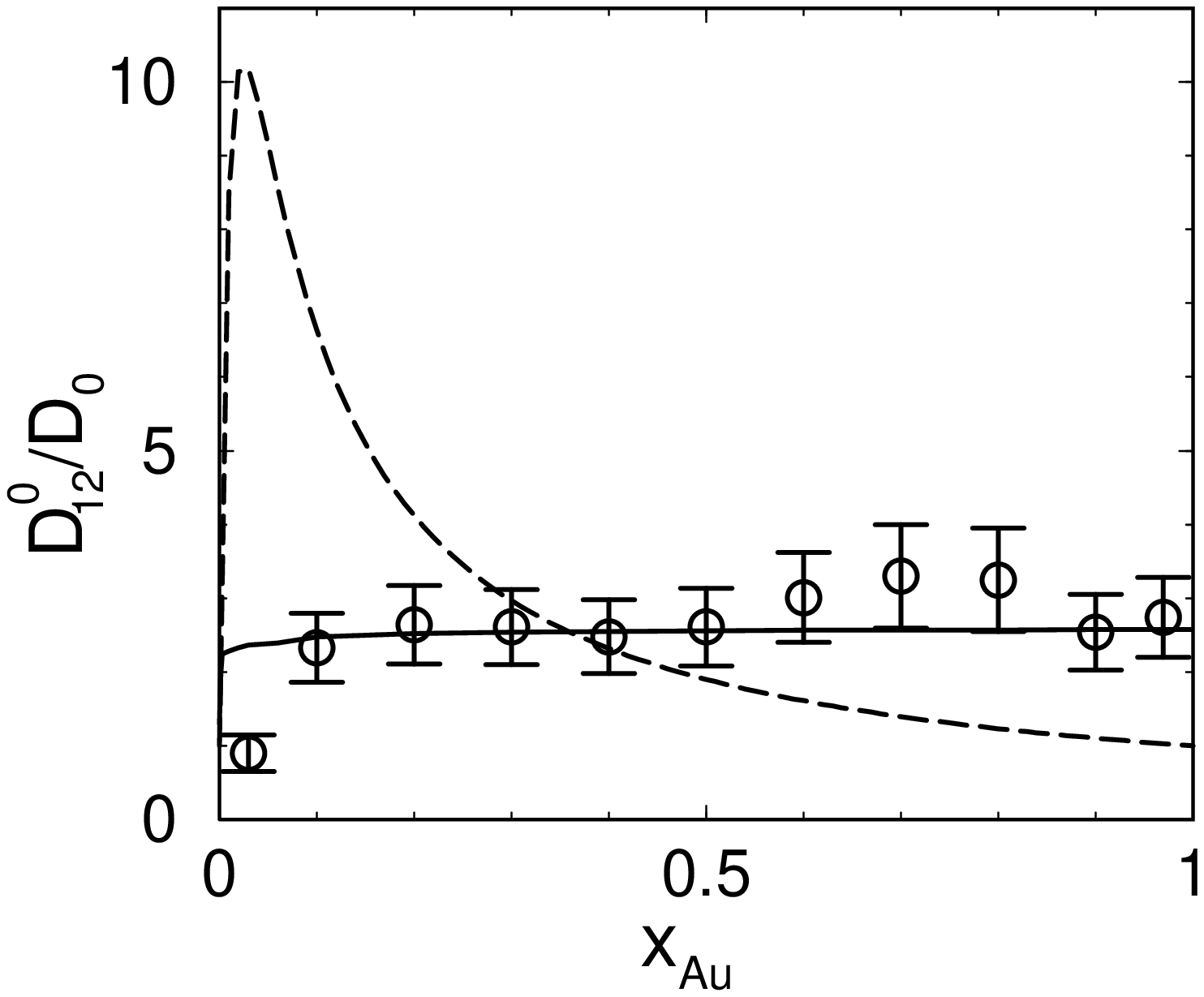}
\caption{Mutual diffusion constant contribution $D^0_{12}$ (see text) of the $D^+-Au^{39+}$ BIM 
at $\Gamma=0.05 $ as a function of composition: simulations (circles), 
screened potential model of Ref. \cite{ppf86} (solid line). Thermodynamic prefactor  
$\langle Z^2\rangle/\langle Z\rangle^2$ (dashed line).}
\label{fig:d12}
\end{figure}

In the course of the molecular dynamics simulations with various $D^+ - Au^{39+}$ 
mixtures we also calculated the microscopic interdiffusion 
current ${\bf j_c}$, and therefore were able to determine the Green-Kubo integrand $D^0_{12}$. 
The results are shown in Fig. \ref{fig:d12} relative to $D_0=a^2\omega_p$, along with the discussed 
thermodynamic prefactor estimate. We find that $D^0_{12}$ is almost concentration independent 
for $x_{Au}\ge 0.1$, but appears to decrease fairly steeply at lower concentrations. The prefactor 
$\langle Z^2\rangle/\langle Z\rangle^2$ has a simple behavior, with a sharp 
maximum for small amounts of highly charged ions \cite{ogata}. 

For a BIM with small ionic asymmetry kinetic theory estimates of the $D^0_{12}$ were found to be in good agreement 
with simulations \cite{bp87}. A simpler model, widely employed for astrophysics problems, was proposed 
by Paquette 
et al. \cite{ppf86}. Its main assumption is that the Boltzmann equation can be used to calculate the transport 
coefficients of plasma mixtures modeled as BIM by making use of the Chapman-Enskog solution method \cite{cc}. 
The authors further argue that screened potentials, $Z_iZ_je^2exp(-r/\lambda)/r$, 
are better suited for such estimates than the pure 
Coulomb interaction. In order to extend the validity of this approach to strong couplings 
they propose as appropriate screening distance $\lambda$ the larger of $\lambda_D$ and $a$, 
where $\lambda_D$ is the Debye screening length:
\begin{eqnarray} 
\lambda_D=\left[\frac{k_B T}{4\pi e^2 (\rho^\prime + \sum_i \rho_i Z^2_i)}\right]^{\frac{1}{2}}
\end{eqnarray}
Under these assumptions the Boltzmann equation mutual diffusion coefficient is given in the first 
Enskog approximation as:
\begin{eqnarray}
[D_{12}]_1=\frac{3k_B T (M_1 + M_2)}{16\rho M_1 M_2 \Omega^{(11)}_{12}}
\end{eqnarray}
where $\Omega^{(11)}_{12}$ are collision integrals \cite{cc} that have been tabulated with high 
accuracy in Ref. \cite{ppf86}. We perform such calculations for the $D^+-Au^{39+}$ mixture 
using the slightly better second Enskog approximation \cite{ppf86}. The outcome, see Fig. \ref{fig:d12}, 
reproduces rather well the MD simulation results for $D^0_{12}$ at $x_{Au}\ge 0.1$, but 
{\it not} the mutual diffusion coefficient $D_{12}$. This is an important point that merits further 
discussion. In fact, there is no reason to expect that Chapman-Enskog estimates based on the 
Boltzmann equation for the ions can reproduce the full $D_{12}$ for either pure Coulomb interactions 
(with some reasonable cut-off) or screened potentials. As shown in \cite{bp87} for low density 
plasmas or more generally here, 
the prefactor value $\langle Z^2\rangle/\langle Z\rangle^2$ 
only arises when the electronic background is explicitly taken into account either through its 
ambipolar field in a dilute plasma kinetic description \cite{bp87} or simply at the thermodynamic level in the 
context of the linear mixing rule. 
No such effect is included when the standard Boltzmann equation, which is consistent with 
ideal gas thermodynamics, is used to model the dynamics of the ions. Therefore it is 
reasonable to expect that such approaches can only provide estimates of the 
Green-Kubo part, i.e. $D^0_{12}$, of the mutual diffusion constant, as already evidenced 
by our simulation results and perhaps not fully appreciated before. We note however that 
the prefactor value quoted above may be a good approximation only for 
$r_s\rightarrow0$. For real systems, e.g. those encountered in ICF or astrophysics problems, 
$r_s$ departs significantly from zero and the electronic pressure and polarization effects 
can lead to phase separation \cite{htv77,ld88}, especially at high 
asymmetries. The thermodynamic quantity $(\partial^2(\beta G/N)/\partial x^2_1)_{P,T}$
is connected to the spinodal decomposition line of the plasma mixture \cite{ld88}, and 
therefore proper estimates require careful calculations. In particular, it was pointed out in 
\cite{ld88} that it is not sufficient to consider a BIM with compressible but non-polarizable
electronic background since a consistent treatment can only be achieved with an appropriate 
modeling of electronic screening.

The apparent failure at low $Au$ concentrations of the screened potentials method proposed by 
Paquette et al. may appear at first puzzling since as $x_{Au}$ decreases so 
does the effective coupling constant $\Gamma_{eff}$, and therefore the accuracy of the theory should 
in principle only increase. We note however that this also requires $\lambda_D > a$, which 
for our system and chosen $\Gamma$ means $x_{Au}$ no bigger 
than $\approx 10^{-3}$. In fact, although 
not easily seen in Fig. \ref{fig:d12}, the theoretical values drop sharply for such compositions
to values close to the MD result at the lowest simulated $Au$ concentration, $x_{Au} = 0.03$.
For this composition the screening distance $a$, although larger than $\lambda_D$, appears to 
be too small. This is perhaps not unexpected given that due to the significant separation between the 
highly-charged ions they are primarily screened by the small ions and the electronic background, which 
requires distances significantly larger than $a$ when the charge asymmetry is very large.

In summary, we have investigated using molecular dynamics simulations the viscosity and mutual diffusion 
coefficients of a plasma model known as the binary ionic mixture (BIM) when the asymmetry is very strong. 
We discuss in light of the MD results important shortcomings of widely used theoretical models at large 
asymmetries. For viscosity, an OCP-based ``one-fluid'' theory proves inadequate for quantitative 
predictions and we propose a new method based on the linear mixing rule. A commonly employed model of 
ionic interdiffusion based on the Boltzmann equation compares reasonable well with the simulation results, 
but we point out that it provides only part of the mutual diffusion coefficient. The missing thermodynamic 
piece may be particularly important for large asymmetries and low concentrations of highly-charged ions, 
situations often encountered in both ICF and astrophysics applications. Finally, calculations that 
take into account electronic polarization effects are currently under way and will be reported in a future 
publication. 

I thank Harry Robey, Peter Amendt and Jose Milovich for informative discussions. 
This work was performed under the auspices of the U. S. Department of Energy by 
University of California Lawrence Livermore National Laboratory under Contract 
No. W-7405-Eng-48.

\newpage
\newpage
\newpage
\newpage


\begin{thebibliography}{99}
\bibitem{ogata}The qualitative behavior of the thermodynamic prefactor remains unchanged 
(but its values are even higher) if deviations from the linear mixing rule are 
taken into account and estimated using the relation proposed in Ref. \cite{oii93} 
(the present charge assymmetry is however larger than the ones studied in 
\cite{oii93}).
\end{thebibliography}
\end{document}